\documentclass[twocolumn,aps,prl,amssymb,amsmath,amsfonts,superscriptaddress,showpacs]{revtex4}
\usepackage{graphicx}
\usepackage[usenames,dvipsnames]{xcolor}
\usepackage{mathrsfs,ulem,epstopdf}
\usepackage{bm}
\usepackage{wasysym}
\usepackage{color}
\def\be{\begin{equation}}
\def\ee{\end{equation}}
\def\bc{\begin{center}}
\def\ec{\end{center}}

\newcommand{\bcso}{BaCuSi$_2$O$_6$}

\newcommand{\bea}{\begin{eqnarray}}
\newcommand{\eea}{\end{eqnarray}}
\usepackage[colorlinks,bookmarks=false,citecolor=red,linkcolor=red,urlcolor=NavyBlue]{hyperref}
\begin{document}
\title{Spatially Resolved Magnetization in the Bose-Einstein Condensed State of {\bcso}:\\ Evidence for Imperfect Frustration}
\author{S. Kr\"amer}
\email{steffen.kramer@lncmi.cnrs.fr}
\affiliation{Laboratoire National des Champs Magn\'etiques Intenses, LNCMI - CNRS (UPR3228),
UJF, UPS and INSA, BP 166, 38042 Grenoble Cedex 9, France}
\author{N. Laflorencie}
\email{laflo@irsamc.ups-tlse.fr}
\affiliation{Laboratoire de Physique Th\'eorique, Universit\'e de Toulouse, UPS, (IRSAMC), 31062 Toulouse, France}
\author{R. Stern}
\affiliation{National Institute of Chemical Physics and Biophysics,  Akadeemia tee 23, 12618 Tallinn, Estonia}
\author{M. Horvati\'c}
\affiliation{Laboratoire National des Champs Magn\'etiques Intenses, LNCMI - CNRS (UPR3228),
UJF, UPS and INSA, BP 166, 38042 Grenoble Cedex 9, France}
\author{C. Berthier}
\affiliation{Laboratoire National des Champs Magn\'etiques Intenses, LNCMI - CNRS (UPR3228),
UJF, UPS and INSA, BP 166, 38042 Grenoble Cedex 9, France}
\author{H. Nakamura}
\affiliation{Division of Materials Physics, Graduate School of Engineering Science, Osaka University, Toyonaka, Osaka 560-8531, Japan}
\author{T. Kimura}
\affiliation{Division of Materials Physics, Graduate School of Engineering Science, Osaka University, Toyonaka, Osaka 560-8531, Japan}
\author{F. Mila}
\affiliation{Institut de Th\'eorie des Ph\'enom\`enes Physiques, \'Ecole Polytechnique F\'ed\'erale de Lausanne (EPFL), CH-1015 Lausanne, Switzerland}

\date{\today}

\begin{abstract}
In order to understand the nature of the two-dimensional Bose-Einstein condensed (BEC) phase in \bcso, we performed detailed $^{63}$Cu and $^{29}$Si NMR above the critical magnetic field, $H_{c1} = 23.4$~T. The two different alternating layers present in the system have very different local magnetizations close to $H_{c1}$; one is very weak, and its size and field dependence are highly sensitive to the nature of inter-layer coupling. Its precise value could only be determined by ``on-site'' $^{63}$Cu NMR, and the data are fully reproduced by a model of interacting hard-core bosons in which the perfect frustration associated to tetragonal symmetry is slightly lifted, leading to the conclusion that the population of the less populated layers is not fully incoherent but must be partially condensed.
\end{abstract}


\maketitle

%

\begin{figure}[!hb]
\centering
\includegraphics[width=0.96\columnwidth,clip]{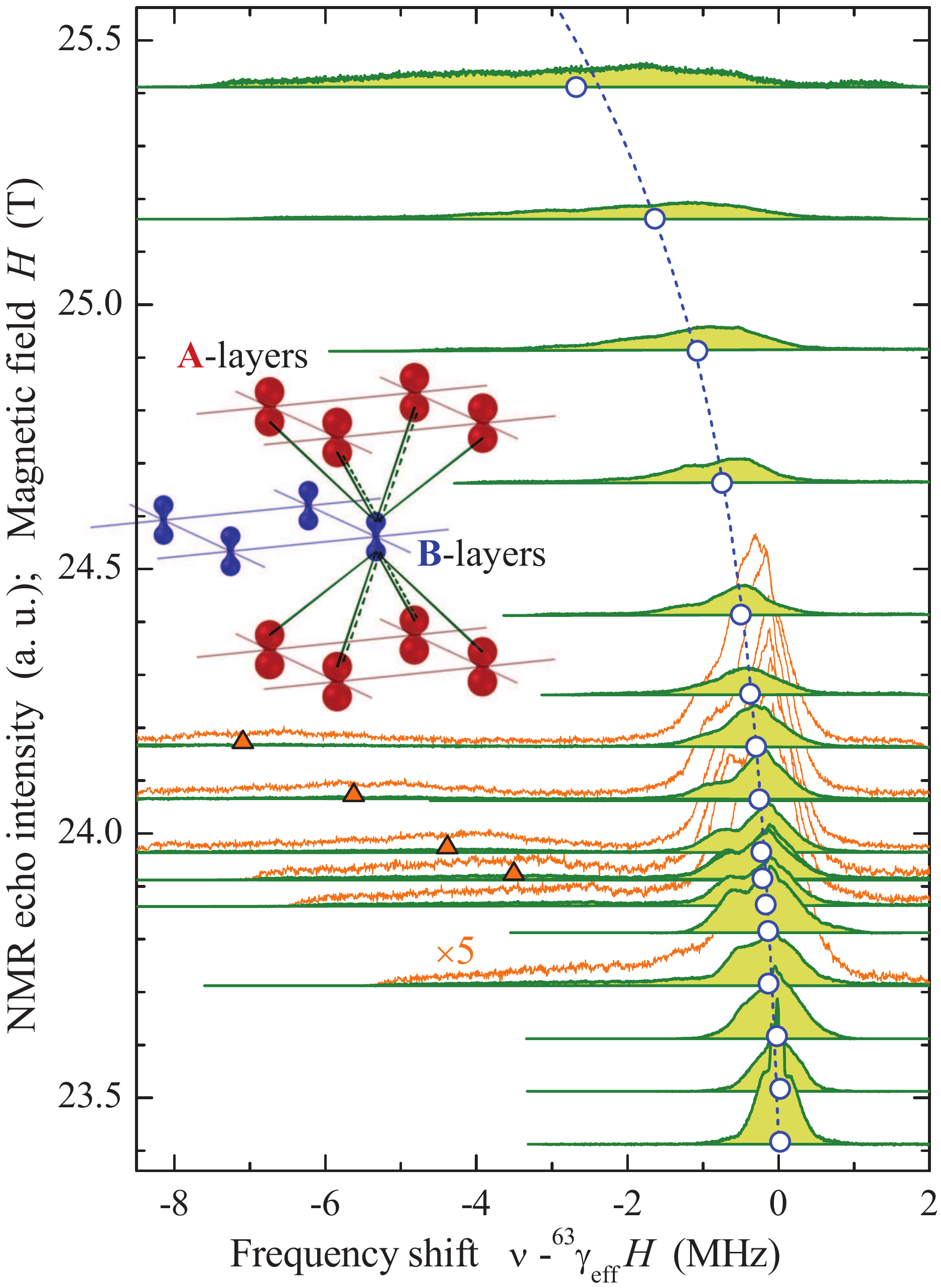}
\caption{(color online) Magnetic field dependence of the low-temperature (0.05-0.08\,K) $^{63}$Cu NMR spectra on entering the BEC phase of BaCuSi$_2$O$_6$ (green lines). The vertical offset of the spectra defines the corresponding field value. The spectra are almost exclusively due to B layers, and their first moment (open circles) measures the corresponding local magnetization. The dotted line is the theoretical prediction, see the text and Fig.~\ref{fig:2}. The $\times$5 zoom of selected spectra (orange lines) reveals a weak, broad and strongly shifted contribution of A layers, and the orange triangles denote a rough estimate of its first moment. Inset shows schematic structure of the spin 1/2 dimers: green lines present perfectly frustrated, symmetric inter-layer exchange couplings, and dotted lines denote weak breaking of this symmetry.}
\label{fig:1}
\end{figure}

\begin{figure}
\centering
\includegraphics[width=\columnwidth,clip]{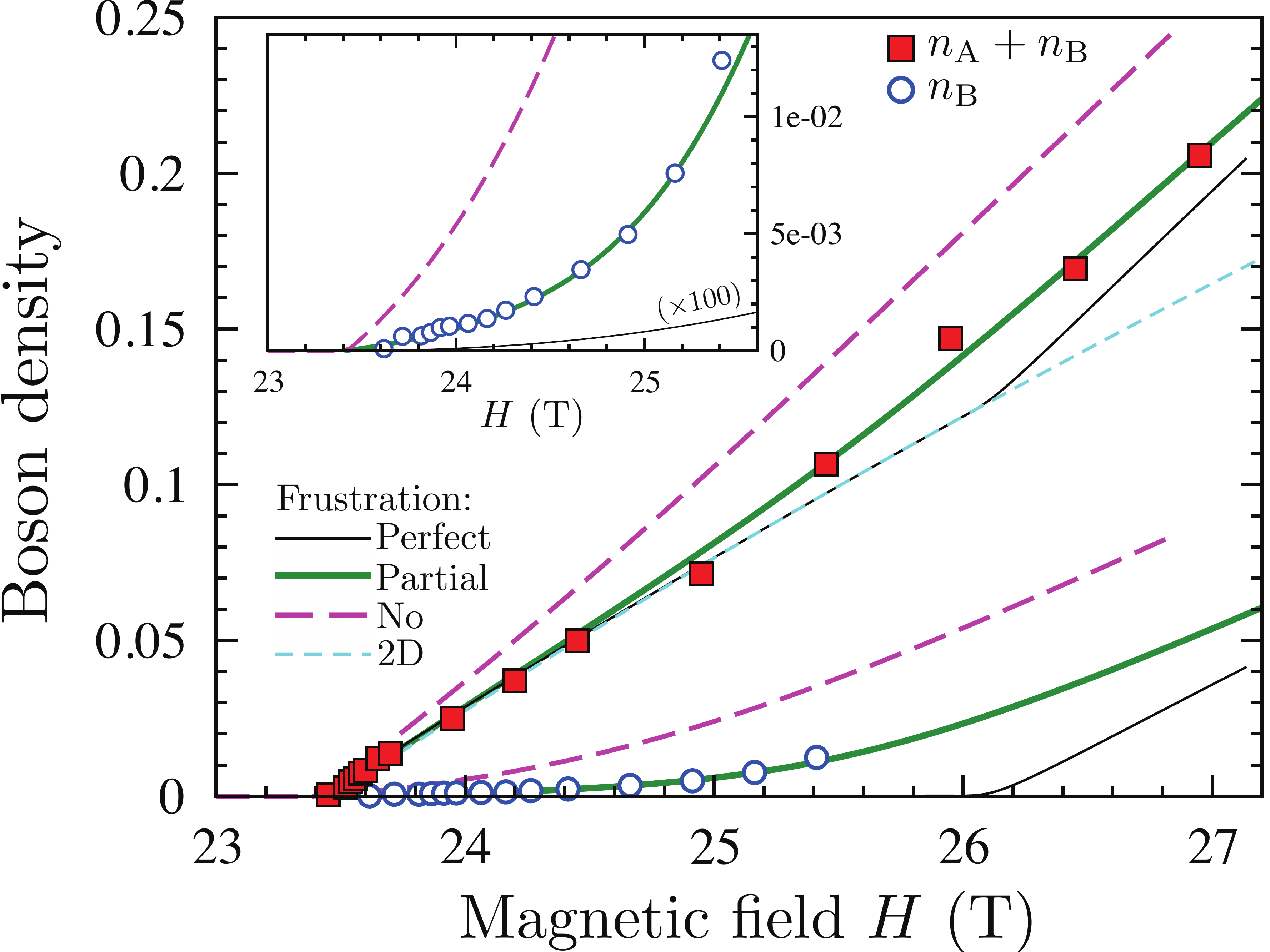}
\caption{(color online) Magnetic field dependence of boson densities. The $n_{\rm B}$ from $^{63}$Cu NMR (open blue circles) and $n_{\rm A}+n_{\rm B}$ from $^{29}$Si NMR (red squares) are compared to three different theoretical models (see the text) having perfect, partial and no frustration (thin black, thick green, and dashed magenta lines, respectively). Thin cyan dashed line shows the boson density of identical, decoupled layers (2D). Inset: Zoom on $n_{\rm B}$ in the vicinity of $H_{c1}$.}
\label{fig:2}
\end{figure}

Weakly coupled antiferromagnetic (AF) spin-1/2 dimers are currently of great interest, since they offer an ideal playground to study the physics of hard core bosons on a lattice~\cite{Giamarchi08}. All these systems have in common a collective singlet ground state, separated by an energy gap from a band of triplet excitations. For magnetic fields $H$ larger than the critical field $H_{c1}$ closing this gap, the dimers acquire a triplet, i.e. a hard-core boson density controlled by the bare chemical potential, that is the reduced magnetic field $h=H - H_{c1}$. Depending on the geometry of the inter-dimer couplings, a rich variety of ground states can occur: canted XY AF order which can be described as Bose-Einstein condensation (BEC) \cite{Giamarchi99,Nikuni00}, observed in TlCuCl$_3$~\cite{Nikuni00,Ruegg03}, \bcso~\cite{Jaime04}, DTN~\cite{Zapf06}, and BPCB~\cite{Klanjsek08}; magnetization plateaus described as Mott-insulators, observed in SrCu$_2$(BO$_3$)$_2$~\cite{Kodama02}; Bose glasses observed in doped DTN~\cite{Yu12} and other systems~\cite{Hong10}; and even supersolids~\cite{SS} for which a physical realization has not been found yet. The hallmark of a BEC system is the field dependence of the temperature $T_{\rm BEC}$ of transition into the ordered phase of canted XY AF type, $T_{\rm BEC}\propto h^\phi$, where $\phi=2/d=2/3$ in the standard 3D case. In 2D, the BEC can only occur at $T=0$.
However, depending on the strength and the nature of the inter-layer coupling which stabilizes the 3D order, the $T_{\rm BEC}$ vs. $h$ dependence of quasi-2D systems can be dominated by the properties of the Quantum Critical Point at $T=0=h$, and exhibit 2D exponent $\phi=1$.\\
\indent During the last few years \bcso, also known as Han purple, has emerged as an outstanding candidate for studying this quasi-2D case. Its $T_{\rm BEC}$ was found to vary linearly with $h$, i.e. $\phi=1$~\cite{Sebastian06,Kramer07}, pointing to a quasi-2D BEC. This was originally attributed to frustration between the Bose-Einstein condensates of equivalent adjacent layers, because the Cu dimers form a body-centered tetragonal (bct) lattice~\cite{Sebastian06,Batista07} (inset of Fig.~\ref{fig:1}). In that scenario, the field dependence of $T_c$ is exotic, linear instead of $h^{2/3}$, but the low temperature phase is a standard 3D condensate. Whether the frustration is perfect or not will not affect the nature of the low temperature phase but simply induce a crossover to the regular $h^{2/3}$ behavior at very low temperature~\cite{Rosch07}. \\
\indent However, \bcso~undergoes around 90 K a first order structural transition~\cite{Samulon06} which introduces some structural incommensurability and, more importantly, two types of alternating layers, A and B, along the $c$ axis. The A layers have a smaller intra-dimer exchange coupling, hence smaller local gap and critical field values, than the B layers, as shown by inelastic neutron scattering~\cite{Ruegg07} and by NMR~\cite{Kramer07}. In that case, whether frustration is perfect or not makes a qualitative difference regarding the low temperature phase: For perfect frustration the B layers are only populated by a frustrated proximity effect from the A layers, and they remain uncondensed up to a second critical field~\cite{Rosch07,Laflorencie09}, with however algebraic correlations at zero temperature~\cite{Laflorencie11}. By contrast, for imperfect frustration the B layers are also populated by a regular proximity effect, and they contain a condensate which, if large enough, dominates their population. It is thus of crucial importance to determine whether the structural distortion leads to a negligible or to a significant departure from perfect frustration, in order to unambiguously identify the nature of the low temperature phase of \bcso~ above the first critical field.\\
\indent A direct way to answer that question is to measure very precisely the \textit{local} magnetization of the \textit{less} polarized B layers. Indeed, as shown in Refs.~\onlinecite{Rosch07,Laflorencie09}, the magnetization, i.e. the boson density, of the less polarized  layers just above $H_{c1}$ depends crucially on the level of frustration: it is expected to grow quadratically with $h$ if the frustration is perfect \cite{Laflorencie09}, and linearly otherwise \cite{Rosch07}. In that respect, the $^{29}$Si NMR data of Ref.~\onlinecite{Kramer07}, which have clearly demonstrated the existence of two types of layers, are not sufficiently precise, mostly because the lines of the two types of layers overlap.\\
\indent In this Letter, we show that this difficulty can be overcome by turning to technically much more difficult $^{63}$Cu NMR measurements in the vicinity of the quantum critical field $H_{c1} = 23.4$ T of \bcso~\cite{Kramer07}. Unlike the $^{29}$Si NMR spectra~\cite{Kramer07}, the $^{63}$Cu NMR signal from the B layers is clearly separated from the one from the A layers, and provides an accurate determination of boson densities in the B layers as a function of $h$ (Fig.~\ref{fig:1}). As predicted in~\cite{Laflorencie09}, a quadratic $h$ dependence has been observed, however, its size is much stronger than the predicted one, and there is also a non-negligible linear component. We attribute this latter to a deviation from the perfect frustration due to the incommensurate (IC) lattice modulation~\cite{Samulon06}, and a weak breaking of the tetragonal symmetry~\cite{Sheptyakov12}, both of which characterize the structural distortion at 90 K.\\
\indent \bcso~can be described as Cu$^{2+}$ spin-1/2 dimers located on a bct lattice with Cu-Cu bonds perpendicular to the $ab$-plane (inset of Fig.~\ref{fig:1}). The real space group at room temperature is indeed $I4_1/acd$~\cite{Sparta04}. However, below 90 K a first order structural transition renders adjacent layers inequivalent~\cite{Sheptyakov12}, and also induces an IC distortion of the lattice along the $b$-axis~\cite{Ruegg07}. This IC distortion converts the typical splitting of each NMR line below $T_{\rm BEC}$~\cite{Klanjsek08} into a continuously distributed spectrum~\cite{Kramer07}.
In addition, although the boson densities $n_{\rm A}$ and $n_{\rm B}$ in the A and B layers are very different, the $^{29}$Si NMR lines originating from the two types of layers overlap due to the small sensitivity of the $^{29}$Si NMR to local boson density. Finally, through a significant long-range (dipolar) coupling, the $^{29}$Si nuclei from the B layers are also somewhat sensitive to the A layer boson density. Altogether, this renders an accurate determination of the small $n_{\rm B}$ values by $^{29}$Si NMR quite difficult. Although experimentally more complicated, a study by $^{63}$Cu NMR presents major advantages. The $^{63}$Cu hyperfine tensor is much larger, $^{63}A_{zz}=-16.4$ T$/\mu_B$~\cite{Kramer07} (defined later, see Eq.~\ref{eq:nAB}), enhancing the sensitivity of $^{63}$Cu NMR line position to small values and variations of local boson density. One thus expects that above $H_{c1}$ the different $n_{\rm A}$ and $n_{\rm B}$ values correspond to separated NMR lines, rendering the determination of $n_{\rm B}$ much easier.\\
\indent  The $^{63}$Cu and $^{29}$Si NMR have been performed in the 23-27~T field range, in a resistive magnet at LNCMI. The sample was a 2.4$\times$2.4$\times$1.2\,mm$^3$ single crystal of \bcso, grown in a mirror furnace, and fully enriched in $^{29}$Si to enhance the $^{29}$Si NMR signal. It was placed in the mixing chamber of a $^3$He-$^4$He dilution refrigerator, with its $c$ axis oriented parallel to the external field $H$, i.e., $c \parallel z$. NMR spectra have been recorded at fixed field by summing up Fourier transforms of the NMR echoes taken at regular frequency steps. The frequency $\nu$ for the $^{63}$Cu NMR spectra corresponds to the central line of $^{63}$Cu, and is taken relative to the reference $\nu_0=~^{63}\gamma_{\rm eff}H$, where $^{63}\gamma_{\rm eff} =~^{63}\gamma(1+ K_{\rm orb})$, with $^{63}\gamma=11.285$ MHz/T and $K_{\rm orb} = 1.68~\%$~\cite{Kramer07}. The value of the applied field $H$ was calibrated with the metallic $^{27}$Al reference placed in the same coil as the sample. \\
\indent In a Bose-Einstein condensate spins (\textbf{\textit{S}}) bear both a longitudinal $M_z = g_z \mu_B\langle S_z\rangle$ and a transverse, staggered component of the magnetization $M_\perp= g_\perp\mu_B\langle S_\perp\rangle$, where $g_{z}$ and $g_{\perp}$ are the corresponding components of the $g$ tensor. In general, both components define the observed NMR frequency through the corresponding components of the hyperfine coupling tensor $A_{zz}$ and $A_{z\perp}$:
\be
\nu_{\pm}=\nu_0+\gamma A_{zz}M_z\pm \gamma A_{z\perp}M_\perp .
\ee
While for the $^{29}$Si NMR in \bcso~this is indeed the case, the hyperfine tensor of the Cu is dominated by on-site interactions (core polarization and on-site dipolar interaction) and is diagonal with its main component $A_{zz}$ along the $c$-axis. The Cu line shapes are thus insensitive to the staggered magnetization (as long as $H \parallel c$) and the frequency shift of the first moment $\nu_1$ (average frequency) of the line, $\Delta\nu=\nu_1-\nu_0$, reflects directly the longitudinal magnetization only, that is the boson density $n_{\rm A(B)}$ by:
\be
n_{\rm A(B)} \equiv 2 \langle S_{z}^{\rm A(B)}\rangle = 2\Delta\nu_{\rm A(B)} / ({^{63}\gamma}g_{cc}\mu_B~ ^{63}A_{zz}),
\label{eq:nAB}
\ee
where $g_{cc}=2.30$~\cite{Zvyagin06}. This formula implies that $n_{\rm A(B)} = 1$ when there is one boson per dimer in the A(B) layer. With this convention, one has $n_{\rm A}+ n_{\rm B} = 2$ at the saturation field $H_{c2} \approx 50$ T. Fig.~\ref{fig:1} presents the field dependence of the $^{63}$Cu NMR line in the BEC phase. The fact that the boson density as well as its fluctuations are much bigger in the A layer makes their contribution to the spectra nearly unobservable for two reasons. First, the intensity of the A line is much smaller, due to a much shorter spin-spin relaxation time $T_2$ reflecting the fluctuations. This mechanism is particularly effective close to the phase transition. Second, the width of the line, being related to boson density, is expected to be much broader for the A line and thus of proportionally smaller intensity (as the total signal is given by the integral over the line). Indeed, in the enhanced scale spectra in Fig.~\ref{fig:1}, the A line can be observed as a very weak broad signal on the left of the main B line only above 23.9 T where the two lines become fully separated. Note that the main line corresponding to the B layer is obviously shifting much more slowly than the one corresponding to the A layer, reflecting a much lower boson density $n_{\rm B}$.\\
\indent Using Eq.~\eqref{eq:nAB}, the first moments of the lines shown in Fig.~\ref{fig:1} are converted to $n_{\rm B}$ and their field dependence is plotted in Fig.~\ref{fig:2}. As $n_{\rm A}$ determined from $^{63}$Cu NMR suffers from huge error bars due to a very poor signal to noise ratio of the A lines, we have instead plotted the total boson density $n_{\rm A} + n_{\rm B}$, deduced from the first moment of the $^{29}$Si line-shapes recorded on the same sample~\cite{Kramer12}. As already mentioned, the contributions of the A and B layers overlap in the $^{29}$Si spectrum, and the lineshape is sensitive to both the transverse magnetization and IC distortion~\cite{Kramer07}. Fortunately, both latter effects average to zero when calculating the first moment $\nu_1$. Therefore, the ${^{29}\nu_1}$ of the total Si spectrum, including the A and B layer contributions, gives an accurate determination of $n_{\rm A} + n_{\rm B}$ through the formula: $n_{\rm A} + n_{\rm B} = 4({^{29}\nu_1}-{^{29}\nu_0})/({^{29}\gamma} g_{cc}\mu_B~{^{29}A_{zz}})$, which is the same formula as Eq.~\eqref{eq:nAB} corrected for a factor of 2 for proper normalization. A possible source of error is the difference in the $^{29}A_{zz}$ values for the A and B layer; here we have used the former one (0.30~T/$\mu_B$) to ensure that the error is negligible in the relevant field range where $n_{\rm B}\ll n_{\rm A}$~\cite{Kramer12}.\\
\indent We now focus on the variation of $n_{\rm B}$ as a function of $h$. In earlier models~\cite{Batista07} all the layers were taken to be identical, which is not realized in \bcso.
When the finite energy barrier, that is the difference between the two local gap values, $\Delta=\Delta_{\rm B}-\Delta_{\rm A} \approx 0.4$ meV~\cite{Ruegg07,Kramer07} is considered, perfect frustration along the $c$-axis leads to a classical decoupling of the layers, and to the \textit{absence} of bosons in B layers close to $H_{c1}$~\cite{Rosch07}, a result in contradiction with our previous~\cite{Kramer07} and new $^{29}$Si NMR data~\cite{Kramer12}, the latter ones being more accurate since recorded on a $^{29}$Si enriched sample. However, as shown in Refs.~\onlinecite{Laflorencie09,Laflorencie11}, quantum fluctuations (spin-wave corrections) above such a classical ground-state induce above $H_{c1}$ a small boson density $n_{\rm B}\propto h^2$. As a corollary, a full three dimensional coherence is restored in this case through an effective inter-layer hopping of bosons, $t_{\rm 3D}\propto h$, leading to a modified exponent $T_{\rm BEC}\propto h^{\phi=1}$, in agreement with experiments~\cite{Sebastian06,Kramer07}.\\
\indent In the inset of Fig.~\ref{fig:2} we plot in detail $n_{\rm B}(h)$, which is an order of magnitude smaller than $n_{\rm A}$, and whose $h$-dependence obviously exhibits a quadratic contribution. However, the measured density $n_{\rm B}$ turns out to be much larger than earlier theoretical predictions based on a perfectly frustrated scenario~\cite{Laflorencie09,Laflorencie11}. Moreover, a linear component is also present in the observed $n_{\rm B}(h)$ dependence. As already mentioned, the key to explain these deviations is an imperfect frustration between A and B layers, which has two origins. According to the recent low-temperature average structure of \bcso~\cite{Sheptyakov12}, the tetragonal symmetry is broken, which introduces two slightly different Cu$_{\rm A}$-Cu$_{\rm B}$ distances, corresponding to the two type of Cu$_{\rm A}$ sites having opposite staggered magnetization. This provides a direct and site-independent lifting of the perfect frustration. Another source of imperfect frustration is the IC modulation in the $b$ direction, which is along the diagonal connecting the next nearest neighboring dimers~\cite{Samulon06}. While both effects are very difficult to estimate, the NMR data provide, with the help of the following theoretical analysis, a reliable estimate of their total ``effective'' size.\\
\indent As discussed previously in Refs.~\onlinecite{Jaime04,Batista07,Laflorencie09,Laflorencie11}, one can map the system of coupled spin-1/2 dimers in BaCuSi$_2$O$_6$ onto a lattice model of hard-core bosons whose quantum dynamics is governed in each layer $\ell=$ A, B by the Hamiltonian
\bea
{{\cal H}}_{\parallel}^{\ell}&=&t_{\parallel}\sum_{
\begin{picture}(12,12)(0,0)
        \put (0,2) {\line (1,0) {12}}
        \put (2,0) {\line (0,1) {12}}
        \put (0,10) {\line (1,0) {12}}
        \put (10,0) {\line (0,1) {12}}
        \put (2,10) {\circle*{3}}
        \put (10,10) {\circle*{3}}
        \put (2,2) {\circle*{3}}
        \put (10,2) {\circle*{3}}
\end{picture}
}
\left[b^{\dagger}_{\ell}({\bf r})b^{\vphantom{\dagger}}_{\ell}({\bf r'})+{\rm{h.c.}}
+n_{\ell}({\bf r})n_{\ell}({\bf r'})\right]\nonumber\\
&-&\mu_{\ell}\sum_{\bf r}n_{\ell}({\bf r}),
\label{eq:X}
\eea
where the sum runs over the nearest neighbors on a square lattice. It is important to notice the positive (antiferromagnetic) hopping term $t_{\parallel}$ and the distinct chemical potentials on different layers: $\mu_{\rm A}=\mu_{\rm B}+\Delta$. The inter-layer (or ``transverse'') tunnelling term, which mixes bosons from layers A and B, reads:
\bea
{\cal{H}}_{\perp}&=&t_{\perp}\sum_{
\begin{picture}(12,12)(0,0)
        {{\put (1,1) {\line (1,1) {10}}}}
        \put (11,1) {\line (-1,1) {10}}
        \put (6,6) {\circle{2.5}}
        \put (2,10) {\circle*{3.5}}
        \put (10,10) {\circle*{3.5}}
        \put (2,2) {\circle*{3.5}}
        \put (10,2) {\circle*{3.5}}
\end{picture}
} \left[b^{\dagger}_{\rm A}({\bf r})b_{\rm B}^{\vphantom{\dagger}}({\bf r'}) + {\rm{h.c.}}+n_{\rm A}({\bf r})n_{\rm B}({\bf r'})\right]\nonumber\\
&+&
t_{\perp}'\sum_{
\begin{picture}(12,12)(0,0)
        \put (11,1) {\line (-1,1) {10}}
        \put (6,6) {\circle{2.5}}
        \put (2,10) {\circle*{3.5}}
       \put (10,10) {\circle*{3.5}}
        \put (2,2) {\circle*{3.5}}
        \put (10,2) {\circle*{3.5}}
\end{picture}
} \left[b^{\dagger}_{\rm A}({\bf r})b_{\rm B}^{\vphantom{\dagger}}({\bf r'}) + {\rm{h.c.}}+n_{\rm A}({\bf r})n_{\rm B}({\bf r'})\right],
\label{eq:perp}
\eea
where the first term accounts for the perfectly frustrated hopping between A and B, and the second term mimic a direct (unfrustrated) tunneling between neighboring layers, which is due to the lattice distortion \cite{Samulon06,Sheptyakov12}. One can immediately rule out the scenario of identical and perfectly decoupled layers ($\Delta=t_\perp=t_\perp'=0$)~\cite{Batista07} for which the boson density is plotted in Fig.~\ref{fig:2} by thin cyan dashed line. Although this gives an approximately correct description of the total measured density, in particular close to $H_{c1}$, the B layers should simply not exist in this case. A realistic model should include different layers A and B ($\Delta\neq 0$) and an inter-layer couplings between them.
In order to get the best description of NMR data, we fix the parameters $t_\parallel=0.3$ meV and $\Delta=0.3$ meV, and allow $t_\perp$ and $t'_\perp$ to vary. A spin-wave analysis of this effective hard-core bosonic Hamiltonian is performed at the linear $1/S$ order~\cite{Laflorencie09,Coletta12}.
We have examined three possible scenarios: (i) perfect frustration with $t_\perp/t_\parallel=1/8$ and $t'_\perp=0$; (ii) no frustration with $t_\perp=0$ and $t'_\perp/t_\parallel=1/20$; (iii) partial frustration with $t_\perp/t_\parallel=1/8$ and $t'_\perp/t_\parallel=1/50$. In all these cases, the inter-layer barrier $\Delta=0.3$ meV has been chosen in order to give the best description of the densities measured by NMR.  This value appears somewhat smaller than the experimental ones~\cite{Ruegg07,Kramer07}, but we expect that the IC modulations, particularly strong in B layers, naturally lead to a reduction of this internal barrier. We now examine in details the three aforementioned scenarios:\\
(i) While a perfect frustration is particularly interesting since it would induce an exotic boson density in the B layers~\cite{Laflorencie11}, it is unfortunately unable to explain the magnitude of the density $n_{\rm B}$. This is evidenced in the inset of Fig.~\ref{fig:2} where we see that the quadratic contribution coming from quantum fluctuations is extremely small. Moreover, the perfect frustration scenario causes a clear cusp in the field dependence of the total density $n_{\rm A}+n_{\rm B}$ around $H \approx 26$~T, while in the experimental data, one observes only a slight bending.\\
(ii) The completely unfrustrated case requires a more detailed discussion. The present analysis focuses on the quantum critical regime where the bosons are very diluted, but we have to keep in mind the full experimental phase diagram, and in particular the maximum $T_{\rm BEC}\approx 3.8$~K value occurring at $H\approx 37$~T~\cite{Jaime04}, which imposes a transverse tunneling $t'_\perp=t_\parallel$/20.  This value of the direct (unfrustrated) hopping leads to theoretical densities $n_{\rm A}$ and $n_{\rm B}$ much larger than the experimental data. In particular, $n_{\rm B}$ is about five times bigger, and the bending of the $n_{\rm A}+n_{\rm B}$ curve has completely disappeared with such a scenario, as seen in Fig.~\ref{fig:2}.\\
(iii) We are then left with a third scenario of partial frustration, able to qualitatively and quantitatively explain all the features observed in NMR experiments: the small but finite density $n_{\rm B}$ which has a pronounced convexity, and the total density $n_{\rm A}+n_{\rm B}$ which smoothly departs from a 2D behavior above 25 T. The theoretical fit given in Fig.~\ref{fig:2} corresponds to a frustrated coupling $t_\perp/t_\parallel=1/8$, dominantly determined by the total density $n_{\rm A}+n_{\rm B}$, and a much smaller direct hopping $t'_\perp/t_\parallel=1/50$, which controls $n_{\rm B}$. This coupling lifts the perfect frustration by $2 t'_\perp/(4 t_\perp + 2 t'_\perp) \approx 7 \% $. As regards the scaling of $T_{\rm BEC}(h)$, we expect the contribution of such a small direct coupling to restore a true 3D critical regime with an exponent $\phi=2/3$ only below $(t'_\perp)^2/\Delta\approx 10$ mK, which is out of the current experimental range of measured $T_{\rm BEC}(h)$ in \bcso~\cite{Sebastian06,Kramer07}.\\
\indent To summarize, using $^{63}$Cu and $^{29}$Si NMR, we have shown that the population of the less polarized layers of \bcso~ is not purely quadratic in $(H-H_{c1})$, and that it is much larger than predicted assuming perfect frustration. This unambiguously demonstrates the presence of \textit{imperfect} frustration of the inter-layer couplings, with an unfrustrated component of about $\approx$ 7 \% of the total inter-layer coupling strength, and leads to the conclusion that the population of the B layers is not fully incoherent, as it would be if frustration was perfect, but that it must be partially condensed as soon as $H>H_{c1}$. To probe directly this small condensate
fraction is a challenging task that is left for future investigation.\\
\indent We acknowledge J. Marcus and B. M\'enaert (Institut N\'eel, Grenoble) for sample conditioning and characterization, and R. Pankow (LNCMI Grenoble) for technical assistance. Part of this work has been supported by the French ANR projects NEMSICOM and Quapris, the EuroMagNET network under the EU contract No. 228043, the Swiss National Fund and MaNEP, as well as the Estonian Research Council and ETF8440. We thank IDRIS and CALMIP for allocation of CPU time.


\begin{thebibliography}{10}
%
\bibitem{Giamarchi08}
T. Giamarchi, Ch. R\"uegg and O. Tchernyshyov,
\href{http://dx.doi.org/10.1038/nphys893}{Nature Physics {\bf 4}, 198 (2008).}
%
\bibitem{Giamarchi99}
T. Giamarchi and A.M. Tsvelik,
\href{http://dx.doi.org/10.1103/PhysRevB.59.11398}{Phys. Rev B {\bf 59} 11398 (1999)}.
%
\bibitem{Nikuni00}
T. Nikuni, M. Oshikawa, A. Oosawa, and H. Tanaka,
\href{http://dx.doi.org/10.1103/PhysRevLett.84.5868}{Phys. Rev. Lett. {\bf 84}, 5868 (2000)}.
%
\bibitem{Ruegg03}
Ch. R\"uegg, N. Cavadini, A. Furrer, H.-U. G\"ude, K. Kr\"amer, H. Mutka, A. Wildes, K. Habicht, and P. Vorderwisch,
\href{http://dx.doi.org/10.1038/nature01617}{Nature {\bf 423}, 62 (2003)}.
%
\bibitem{Jaime04}
M. Jaime, V. F. Correa, N. Harrison, C. D. Batista, N. Kawashima, Y. Kazuma, G. A. Jorge, R. Stern, I. Heinmaa, S. A. Zvyagin, Y. Sasago, and K. Uchinokura,
\href{http://dx.doi.org/10.1103/PhysRevLett.93.087203}{Phys. Rev. Lett. {\bf 93}, 087203 (2004)}.
%
\bibitem{Zapf06}
V. S. Zapf, D. Zocco, B. R. Hansen, M. Jaime, N. Harrison, C. D. Batista, M. Kenzelmann, C. Niedermayer, A. Lacerda, and A. Paduan-Filho,
\href{http://dx.doi.org/10.1103/PhysRevLett.96.077204}{Phys. Rev. Lett. {\bf 96}, 077204 (2006)};
S. A. Zvyagin, J. Wosnitza, A. K. Kolezhuk, V. S. Zapf, M. Jaime, A. Paduan-Filho, V. N. Glazkov, S. S. Sosin, and A. I. Smirnov,
\href{http://dx.doi.org/10.1103/PhysRevB.77.092413}{Phys. Rev. B {\bf 77}, 092413 (2008)}.
%
\bibitem{Klanjsek08}
M. Klanj\v{s}ek, H. Mayaffre, C. Berthier, M. Horvati\'c, B. Chiari, O. Piovesana, P. Bouillot, C. Kollath, E. Orignac, R. Citro, and T. Giamarchi,
\href{http://dx.doi.org/10.1103/PhysRevLett.101.137207}{Phys. Rev. Lett. {\bf 101}, 137207 (2008)};
Ch. R\"uegg, K. Kiefer, B. Thielemann, D. F. McMorrow, V. Zapf, B. Normand, M. B. Zvonarev, P. Bouillot, C. Kollath, T. Giamarchi, S. Capponi, D. Poilblanc, D. Biner, and K. W. Kr\"amer, \href{http://dx.doi.org/10.1103/PhysRevLett.101.247202}{Phys. Rev. Lett. {\bf 101}, 247202 (2008)}.
%
\bibitem{Kodama02}
H. Kageyama, K. Yoshimura, R. Stern, N. V. Mushnikov, K. Onizuka, M. Kato, K. Kosuge, C. P. Slichter, T. Goto, and Y. Ueda,
\href{http://dx.doi.org/10.1103/PhysRevLett.82.3168}{Phys. Rev. Lett. {\bf 82}, 3168 (1999)};
K. Kodama,  M. Takigawa, M. Horvati\'c, C. Berthier, H. Kageyama, Y. Ueda, S. Miyahara, F. Becca, F. Mila,
\href{http://dx.doi.org/10.1126/science.1075045}{Science {\bf 298}, 395 (2002)}.
%
\bibitem{Yu12}
R. Yu, L. Yin, N. S. Sullivan, J. S. Xia, C. Huan, A. Paduan-Filho, N. F. Oliveira Jr, S. Haas, A. Steppke, C. F. Miclea, F. Weickert, R. Movshovich, E.-D. Mun, B. L. Scott,    V. S. Zapf , T. Roscilde,
\href{http://dx.doi.org/10.1038/nature11406}{Nature {\bf489}, 379 (2012)}.
%
\bibitem{Hong10}
T.~Hong, A.~Zheludev, H.~Manaka, and L.-P. Regnault,
\href{http://dx.doi.org/10.1103/PhysRevB.81.060410}{Phys. Rev. B {\bf 81}, 060410 (2010)}.
%
\bibitem{SS}
K.-K. Ng and T. K. Lee,
\href{http://dx.doi.org/10.1103/PhysRevLett.97.127204}{Phys. Rev. Lett. {\bf 97}, 127204 (2006)};
P. Sengupta and C. D. Batista,
\href{http://dx.doi.org/10.1103/PhysRevLett.98.227201}{Phys. Rev. Lett. {\bf 98}, 227201 (2007)};
N. Laflorencie and F. Mila,
\href{http://dx.doi.org/10.1103/PhysRevLett.99.027202}{Phys. Rev. Lett. {\bf 99}, 027202 (2007)}.
%
\bibitem{Sebastian06}
S. E. Sebastian, N. Harrison, C. D. Batista, L. Balicas, M. Jaime, P. A. Sharma, N. Kawashima, and I. R. Fisher,
\href{http://dx.doi.org/10.1038/nature04732}{Nature {\bf 441}, 617 (2006)}.
%
\bibitem{Kramer07}
S. Kr\"amer, R. Stern, M. Horvati\'c, C. Berthier, T. Kimura, and I. R. Fisher,
\href{http://dx.doi.org/10.1103/PhysRevB.76.100406}{Phys. Rev. B {\bf 76}, 100406(R) (2007)}.
%
\bibitem{Batista07}
C. D. Batista, J. Schmalian, N. Kawashima, P. Sengupta, S. E. Sebastian, N. Harrison, M. Jaime, and I. R. Fisher,
\href{http://dx.doi.org/10.1103/PhysRevLett.98.257201}{Phys. Rev. Lett. {\bf 98}, 257201 (2007)};
J. Schmalian and C. D. Batista,
\href{http://dx.doi.org/10.1103/PhysRevB.77.094406}{Phys. Rev. B {\bf 77}, 094406 (2008)}.

\bibitem{Rosch07}
O. R\"osch and M. Vojta,
\href{http://dx.doi.org/10.1103/PhysRevB.76.180401}{Phys. Rev. B {\bf 76}, 180401(R) (2007)};
O. R\"osch and M. Vojta,
\href{http://dx.doi.org/10.1103/PhysRevB.76.224408}{Phys. Rev. B {\bf 76}, 224408 (2007)}.
%
\bibitem{Samulon06}
E.C. Samulon, Z. Islam, S.E. Sebastian, P.B. Brooks, M.K. McCourt, J. Ilavsky, and I.R. Fisher,
\href{http://dx.doi.org/10.1103/PhysRevB.73.100407}{Phys. Rev. B {\bf 73}, 100407 (2006)}.
%
\bibitem{Ruegg07}
Ch. R\"uegg, D. F. McMorrow, B. Normand, H. M. R{\o}nnow, S. E. Sebastian, I. R. Fisher, C. D. Batista, S. N. Gvasaliya, Ch. Niedermayer, and J. Stahn,
\href{http://dx.doi.org/10.1103/PhysRevLett.98.017202}{Phys. Rev. Lett. {\bf 98}, 017202 (2007).}
%
\bibitem{Laflorencie09}
N. Laflorencie and F. Mila,
\href{http://dx.doi.org/10.1103/PhysRevLett.102.060602}{Phys. Rev. Lett. {\bf 102}, 060602 (2009)}.
%
\bibitem{Laflorencie11}
N.~Laflorencie and F.~Mila,
\href{http://dx.doi.org/10.1103/PhysRevLett.107.037203}{Phys. Rev. Lett. {\bf 107}, 037203 (2011)}.
%
\bibitem{Sheptyakov12}
D. V. Sheptyakov, V. Yu. Pomjakushin, R. Stern, I. Heinmaa, H. Nakamura, and T. Kimura,
\href{http://dx.doi.org/10.1103/PhysRevB.86.014433}{Phys. Rev. B {\bf 86}, 014433 (2012)}.
%
\bibitem{Sparta04}
K.M. Sparta and G. Roth,
\href{http://dx.doi.org/10.1107/S0108768104011644 }{Acta Cryst. B {\bf 60}, 491 (2004)}.
%
\bibitem{Zvyagin06}
S. A. Zvyagin, J. Wosnitza, J. Krzystek, R. Stern, M. Jaime, Y. Sasago, and K. Uchinokura,
\href{http://dx.doi.org/10.1103/PhysRevB.73.094446}{Phys. Rev. B {\bf 73}, 094446 (2006)}.
%
\bibitem{Kramer12}
S. Kr\"amer {\it et al}., unpublished.
%
\bibitem{Coletta12}
T. Coletta, N. Laflorencie and F. Mila,
\href{http://dx.doi.org/10.1103/PhysRevB.85.104421}{Phys. Rev. B {\bf 85}, 104421 (2012)}.
%
\end{thebibliography}
\end{document}